\makeatletter
\declare@file@substitution{revtex4-1.cls}{revtex4-2.cls}
\makeatother
\documentclass[twocolumn, times, tighten,twocolappendix]{aastex631}

\usepackage{graphicx,multirow,hyperref,url,color,xspace}
\usepackage{amsmath,amssymb}

\newcommand{\msun}{{M}_{\sun}}

\newcommand{\nustar}{{\textit{NuSTAR}}\xspace}

\newcommand{\source}{{Swift J1727.8--1613}\xspace}

\newbox\grsign \setbox\grsign=\hbox{$>$} \newdimen\grdimen \grdimen=\ht\grsign
\newbox\simpropbox
\setbox\simpropbox=\hbox{\raise.5ex\hbox{$\propto$}\llap
 {\lower.5ex\hbox{$\sim$}}}\ht2=\grdimen\dp2=0pt
\def\simprop{\mathrel{\copy\simpropbox}}

\defcitealias{Wood24}{W24}
\newcommand{\wood}{{\citetalias{Wood24}}\xspace}
\defcitealias{Wood25}{W25}

\defcitealias{Mata_Sanchez25}{MS25}
\newcommand{\mata}{{\citetalias{Mata_Sanchez25}}\xspace}

\begin{document}

\title{A Novel Method of Modeling Extended Emission of Compact Jets: Application to Swift J1727.8--1613}
\shorttitle{The Jet of Swift J1727.8--1613}

\author[0000-0002-0333-2452]{Andrzej A. Zdziarski}
\affiliation{Nicolaus Copernicus Astronomical Center, Polish Academy of Sciences, Bartycka 18, PL-00-716 Warszawa, Poland; \href{mailto:aaz@camk.edu.pl}{aaz@camk.edu.pl}}

\author[0000-0002-2758-0864]{Callan M. Wood}
\affiliation{International Centre for Radio Astronomy Research, Curtin University, GPO Box U1987, Perth, WA 6845, Australia}

\author[0000-0002-0426-3276]{Francesco Carotenuto}
\affiliation{INAF - Osservatorio Astronomico di Roma, Via Frascati 33, I-00078, Monteporzio Catone, Italy}

\shortauthors{Zdziarski et al.}

\begin{abstract}
Flat radio spectra of compact jets launched by both supermassive and stellar-mass black holes are explained by an interplay of self-absorbed synchrotron emission up to some distance along the jet and optically thin synchrotron at larger distances \citep{BK79}. Their spatial structure is usually studied using core shifts, in which the position of the peak (core) of the emission depends on the frequency. Here, we propose a powerful novel method to fit the spatial dependence of the flux density at a given frequency of the jet and counterjet (when observed) using the theoretical spatial dependencies, which we provide as simple analytical formulae. We apply our method to the spatial structure of the jets in the luminous hard spectral state of the black-hole X-ray binary Swift J1727.8--1613. It was the most resolved continuous jet from an X-ray binary ever observed. We find that the observed approaching jet is significantly intrinsically stronger than the receding one, which we attribute to an increase in the emission of both jets with time (observationally confirmed), together with the light travel effect, causing the receding jet to be observed at an earlier epoch than the approaching one. The jets are relatively slow, with the velocity $\sim(0.3$--$0.4)c$. Our findings imply that the magnetic field strength increased with time. Also, the magnetic flux is much lower than in jets launched by `Magnetically Arrested Disks'. Our method is general, and we propose that it be applied to jets launched by stellar-mass and supermassive black holes.
\end{abstract}

\section{Introduction}
\label{intro}

Compact, or core, extragalactic jets are continuous jets launched by accreting black holes (BHs) in active galactic nuclei (AGNs) and extending up to $\sim\! 10^9 r_{\rm g}$, where $r_{\rm g}=GM/c^2$ is the gravitational radius and $M$ is the BH mass \citep{Hardcastle98, Mullin08}. They have flat radio spectra, with the spectral index of $\alpha\sim 0$, defined by the flux density of $F_\nu \propto \nu^\alpha$, where $\nu$ is the frequency. \citet{Yuan18} considered a large sample of radio-loud AGNs inclined at relatively large angles to the observer (which allowed accurate resolving of their core jets and lobes), and measured the spectra of the core jets only, finding $\langle\alpha\rangle= 0.001$ with $\sigma=0.397$. That type of radio emission is generally accepted as originating from a partially absorbed synchrotron process \citep{BK79, Konigl81}. The emission at a given $\nu$ is synchrotron self-absorbed and hard, with $\alpha=5/2$, up to a distance from the BH of $z_{\rm c}\simprop \nu^{-1}$, and then becomes optically thin and soft, with $\alpha=(1-p)/2$, where $p$ is the power law index of the emitting relativistic electrons. The sum of the emission from the entire jet results in the flat spectrum with $\alpha\approx 0$ up to a break frequency, $\nu_{\rm b}$, above which the entire emitting jet becomes optically thin, with $\alpha=(1-p)/2$. The emission of a compact jet at a given frequency peaks at $z_{\rm c}$, which location is also called a `core'. The core location is offset from that of the BH.

The analogs of extragalactic core jets in BH X-ray binaries (XRBs) are compact jets; see \citet{FBG04} for a review. They also have the radio spectra with $\alpha\sim 0$ and, when possible, are spatially resolved up to $\sim\! 10^9 r_{\rm g}$, e.g., \citet{Stirling01}, and \citet{Wood24}, hereafter \wood. Compact jets are not observed to propagate to large distances.

The spatial structure of both compact XRB jets and extragalactic core jets has been studied with similar methods. First, radio maps are presented (e.g., \citealt{Stirling01} for Cyg X-1 and \citealt{Jorstad05} for a sample of radio loud AGNs). The velocities are measured by studying the motion of jet knots (e.g., \wood and \citealt{Lister19} for jets in XRBs and AGNs, respectively). Third, shifts of the position of the peak or centroid radio emission with frequency are measured. This phenomenon was first used to derive physical properties of extragalactic core jets by \citet{Lobanov98} and later widely used to derive the jet magnetic fields and jet power, e.g., \citet{Zamaninasab14}. Core shifts have also been measured in compact XRB jets, e.g., \citet{Prabu23}, \citet{Wood25}. A method equivalent to core shift is measuring radio time lags between different frequencies, first done by \citet{Tetarenko21} for the XRB MAXI J1820+070, as discussed in section 4.5 of \citet{Zdziarski22a}. Another related approach was that by \citet{Paragi13}, who considered the dependencies between the flux density at a single $\nu$, the core position, and the core size for several observations taken at different times. 

In this paper, we propose a novel method of deriving physical jet parameters. The method utilizes images of compact jets, and it is complementary to the previous methods. Both Galactic and extragalactic compact jets are narrow \citep{Miller-Jones06, Pushkarev09}, and integrating their flux density perpendicularly to the jet direction gives a physical quantity, the flux density per unit distance along the jet. We can compare it with model predictions, in particular with those of \citet{BK79} and \citet{Konigl81}. Here, we use the formulation of that model developed in Appendix A of \citet{Zdziarski19b} and derive an analytical formula for ${\rm d}F_\nu/{\rm d}z$. This formula can be fitted to an observed spatial profile, which then allows us to constrain several physical parameters of jets, in particular, the magnetic field strength and its flux, the rate of the flow of the emitting e$^\pm$, the equipartition and magnetization parameters, and the jet power. The method is general and could be used with jet theoretical models different than those of \citet{BK79} and \citet{Konigl81}. Also, it can be used for both Galactic and extragalactic jets. 

We then apply that method to the compact jet observed by \wood in the luminous hard (or hard intermediate) spectral state of the XRB \source on 2023 August 30. As found by \wood, those observations show the largest resolved compact jet ever seen in an XRB. The results of \wood include an accurate determination of the emission spatial profiles of both the jet and counterjet, which we use.

\source is a transient low-mass XRB (LMXB) whose outburst was first detected on 2023 August 24 by Swift/BAT \citep{Page23}. It was then monitored by MAXI \citep{Matsuoka09}, which showed it reached $\sim$7 Crabs in the 2--20 keV range. The source was also observed by several space and ground-based observatories. The nature of the source as a BH LMXB was firmly established by \citet{Mata_Sanchez25}, hereafter \mata. 

The structure of our paper is as follows. The results specific to \source are given in Sections \ref{parameters} (the source parameters), \ref{data} (the radio data), \ref{results} (results), and Appendix \ref{corner_plot}. The description of our model and method is given in Section \ref{model} and Appendix \ref{jet}. Appendix \ref{HJ88} compares our method to some others in the literature. A discussion of both our new method and its application to \source is given in Section \ref{discussion}, and we conclude in Section \ref{conclusions}. 

\section{The parameters of \source}
\label{parameters}

The distance to the source was estimated as $D=3.4\pm 0.3$ kpc by \mata, while \citet{Burridge25} obtained $D=5.5_{-1.1}^{+1.4}$ kpc by using the absolute red magnitude of a K4V star of $M_r=6.6\pm 0.5$ and the extinction in the red band as $A_r\approx 2.271 E(B$--$V)$ \citep{Schlafly11}. These results are discussed in detail in \citet{Burridge25}, in particular, considering the impact of the observational upper limit of the donor rotational velocity of \mata.

\begin{figure}
\centerline{\includegraphics[width=0.78\columnwidth]{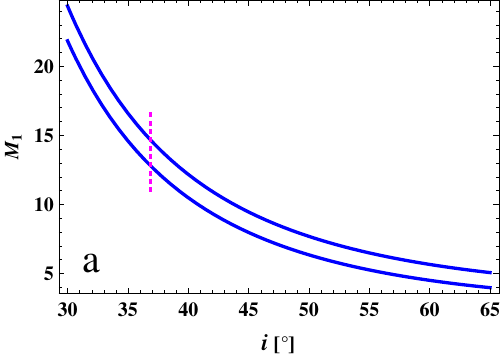}}
\centerline{\includegraphics[width=0.78\columnwidth]{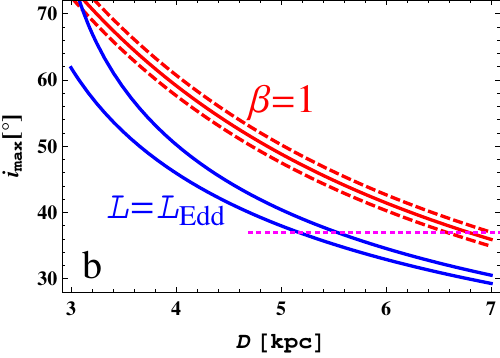}}
\centerline{\includegraphics[width=0.77\columnwidth]{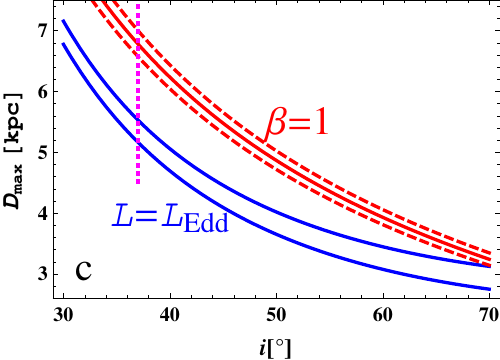}}
\caption{(a) $M_1(i)$ from the mass function assuming the upper limit on $f$ and $M_2=0.78\msun$ and the lower limit on $f$ and $M_2=0.2\msun$, shown by the upper and lower blue curves, respectively. (b) The solid and dashed red curves give the upper limit on $i(D)$ from the angular velocity of the fastest knot observed by \citet{Wood25} for the best fit of $\mu$ and their limits, respectively, see Equation (\ref{i_D}). The blue solid curves give the upper limits corresponding to $L=L_{\rm Edd}$ for the luminous hard state based on \citet{Liu24} with the same assumptions on $M_1$ as in panel (a). (c) The solid and dashed red curves give the upper limit on $D(i)$ from Equation (\ref{i_D}), analogous to panel (b). The blue curves show the upper limits on $D$ assuming $L_{\rm bol}=L_{\rm Edd}$. The dotted magenta lines at $i=37\degr$ on all panels correspond to $M_1\approx 13\msun$.
}
 \label{M_D}
 \end{figure}

\mata obtained the mass function as
\begin{equation}
f\equiv \frac{M_1\sin^3 i}{(1+M_2/M_1)^2}=2.77\pm 0.09\msun,
\label{mass}
\end{equation}
where $M_1$ is the BH mass and $M_2$ is the donor mass, constrained by them as $<0.78\msun$ from the spectral type of the donor of K3--5 V, and we assume $M_2>0.2\msun$ (which mass only weakly affects $i$ and $M_1$). Equation (\ref{mass}) can be solved for $i(M_1,f)$. Figure \ref{M_D}(a) shows then the upper limits of $M_1(i)$ given the uncertainties of $f$ and $M_2$. We note that the highest BH masses found so far in an LMXB by dynamical measurements (see the catalog by \citealt{Fortin24}) are those of GRS 1124--684, which has $M_1=11.0^{+2.1}_{-1.4}\msun$ \citep{Wu16_NM}, and of GRS 1915+105, which has $M_1=11.2^{+1.9}_{-1.7}\msun$ \citep{Reid14, Reid23}. Also, as noted by \mata, \source has almost the same orbital period and the donor type as GRS 1124--684. Thus, a tentative constraint is $M_1\leq 13\msun$, though higher masses are possible. This yields $i\gtrsim 37\degr$, which we show by the dotted magenta lines in Figure \ref{M_D}. 
 
As shown by \wood, the jet position angle is identical within errors to the X-ray polarization angle measured by Imaging X-ray Polarimetry Explorer (IXPE; \citealt{Weisskopf22}), see \citet{Veledina23,Ingram24}. This is compatible with the jet and the binary inclinations being aligned, which we hereafter assume. This polarization angle also implies that the X-ray source is extended perpendicularly to the jet. 

\wood measured the apparent velocity of a prominent southern jet knot present in their VLBA image taken on 2023 September 19--20 as $\mu =0.66\pm 0.05$ mas hr$^{-1}$. Its relation to the velocity, $\beta c$ and $i$ implies
\begin{equation}
\beta=\frac{D\mu}{D\mu\cos i+c \sin i}.
\label{beta_mu}
\end{equation}
Assuming that the core jet velocity equals that of the knot, we obtain $\beta\sim 0.3$--0.6 at $D=3$--7 kpc. However, the knot could have a different velocity than the core jet. This is supported by the observations of three jet knots in a later observation moving with three different apparent velocities \citep{Wood25}. The fastest knot had $\mu =3.18\pm 0.10$ mas hr$^{-1}$. That measurement gives upper limits on the jet inclination vs.\ distance and on the distance vs.\ inclination by setting $\beta=1$ in Equation (\ref{beta_mu}),
\begin{equation}
i_{\rm max}=\arccos\frac{(D\mu/c)^2-1}{(D\mu/c)^2+1}, \quad
D_{\rm max}=\frac{c}{\mu}\cot\frac{i}{2},
\label{i_D}
\end{equation}
respectively. We show the former and latter constraints in Figure \ref{M_D}(b,  c) (the upper curves), respectively. These are hard limits since $\beta<1$ for any source. The Lorentz factors of ejected blobs for other sources span the range of  $\Gamma\sim 2$ to several \citep{Carotenuto24, Matthews25}, which, in turn, would reduce the limit on $D$. \citet{Wood25} gave the limit on $i$ for the distance determination of \citet{Mata_Sanchez25}.  

The jet velocity and $i$ were also constrained by \wood from the jet/counterjet flux ratio, assuming they are intrinsically identical. However, we are showing in the present work that the {\it observed\/} jet is intrinsically stronger than the counterjet, which is possible owing to the temporal evolution and the light travel lag between the two components; see Section \ref{results} below. Therefore, we do not use that constraint.

Another important constraint follows from the fact that the source was very luminous in X-rays in the hard and hard-intermediate spectral states. As shown in figure 1 of \citet{Veledina23}, the 2--20 keV photon flux measured by MAXI reached $>$30 cm$^{-2}$ s$^{-1}$, which is, for example, $\approx$4 times higher than that of another very luminous BH LMXB MAXI J1820+070, whose distance is $\approx$3 kpc \citep{Atri20}, and the BH mass is $6.75^{+0.64}_{-0.46}\msun$ \citep{Mikolajewska22}. \citet{Liu24} fitted broadband X-ray spectra from the luminous hard state of \source. We have calculated the unabsorbed bolometric flux by reproducing their best model (3) for their (brightest) epochs 4 and 5 and found the flux to be as large as $F_{\rm bol}\approx 5\times 10^{-7}$ erg cm$^{-2}$ s$^{-1}$. We note that \citet{Liu24} give their total flux as $\approx 3\times 10^{-7}$ erg cm$^{-2}$ s$^{-1}$, but that number is for the absorbed 2--120 keV only (Yan-Jun Xu, private communication). 

The highest luminosities observed in individual observations of the hard or hard intermediate states satisfy $\lesssim 0.3 L_{\rm Edd}$ \citep{DG03, DGK07}, where $L_{\rm Edd}\equiv 4\pi G M_1 m_{\rm p} c/\sigma_{\rm T}$ is the Eddington luminosity for hydrogen, $m_{\rm p}$ is the proton mass, and $\sigma_{\rm T}$ is the Thomson cross section. Here, we conservatively assume $L_{\rm bol}\leq L_{\rm Edd}$ ($\approx 1.3\times 10^{39}$ erg s$^{-1}$ at $10\msun$), which yields
\begin{equation}
D_{\rm max}= \left(\frac{G M_1(i,M_2) m_{\rm p}}{F_{\rm bol}\sigma_{\rm T}}\right)^{1/2},
\label{Dmax}
\end{equation}
where $M_1(i,M_2)$ is the solution to the cubic Equation (\ref{mass}). The two lower curves in Figures \ref{M_D}(b, c) show the resulting upper limits on $i(D)$ (obtained by solving Equation \ref{Dmax} for $i$) and $D(i)$,  respectively, with their range due to the uncertainties of $f$ and $M_2$, and for $F_{\rm bol}= 5\times 10^{-7}$ erg cm$^{-2}$ s$^{-1}$. The dotted magenta lines correspond to the value of $M_1\approx 13\msun$, at which $D\lesssim 5.1$--5.3 kpc in Figure \ref{M_D}(c). However, the high brightness of \source may be due to the BH mass being $>\! 13\msun$, in which case higher distances would be possible. 

We note that there have been some reported cases of sources with hard spectra with a larger Eddington ratio. Table 15 of \citet{Tetarenko16} lists several cases with the hard-state luminosities $>10^{39}$ erg s$^{-1}$. While such luminosities are reported there, we are not aware of any hard-state spectra showing that, when studied individually. A single case of $L> L_{\rm Edd}$ in the BH LMXB V404 Cyg was observed in its hard flaring state \citep{Zycki99}. While that spectrum was relatively hard, those authors note that the variability properties were significantly different than those typical for the hard state, and \citet{Zycki99} rule out that it belongs to that state. For \source, the power spectrum during its luminous hard (or hard intermediate) state was typical for that state \citep{Ingram24}. 

On the other hand, the source went from the soft state back to the hard one at a very low flux. We estimated $F_{\rm bol}\approx 8\times 10^{-9}$ erg cm$^{-2}$ s$^{-1}$ during 2024 April 3--8 (MJD 60403--60408) after the return to the hard state assuming a Comptonization spectrum with the parameters based on figure 5 of \citet{Podgorny24} and matching their 2--8 keV spectrum from IXPE. We also fitted the Nuclear Spectroscopic Telescope Array (\nustar; \citealt{Harrison13}) observation on 2024 Feb.\ 21 (MJD 60361; Obs\_ID 80902348007), shortly before the transition to the hard state. That spectrum was disk-dominated, and we obtained $F_{\rm bol}\approx 10\times 10^{-9}$ erg cm$^{-2}$ s$^{-1}$, i.e., very similar to the above one.

Thus, the ratio of the maximum hard-state flux to that of the transition is $\approx$60. This is similar to the case of XTE J1550--564, where $L/L_{\rm Edd}\sim 0.2$ at the peak of the hard state and $L/L_{\rm Edd}\sim 0.003$ at the soft-to-hard transition \citep{DG03, DGK07}. While soft-to-hard transitions occur typically at $\approx$0.01--$0.02 L_{\rm Edd}$ \citep{Maccarone03}, these two sources appear different.

\section{The data}
\label{data}

The VLBA observation reported in \wood was done on 2023 August 30 (MJD 60186). The source was in the luminous hard (also classified as hard intermediate) spectral state. The observation followed the first radio detection of \source four days earlier, on 2023 August 25.98 \citep{Miller-Jones23}. 

We denote the angular separation from the BH by $\xi$, and that from the centroid of the image, by $\xi'$, and define $\Delta\xi\equiv \xi-\xi'$. The two quantities differ, as discussed in Section \ref{intro}. The relationship between the physical jet length, $z$, and $\xi$ in units of 1\,milliarcsecond (mas; $\approx 4.85\times 10^{-9}$), is
\begin{equation}
z=\xi z_1,\quad z_1\equiv 1\,{\rm mas}\, D /\sin i,\quad 
\frac{{\rm d}z}{{\rm d}\xi}=z_1,
\label{mas}
\end{equation}
where $i$ is the jet viewing angle. E.g., at the distance of 4 kpc, $1\,{\rm mas}\approx 6.0\times 10^{13}/\sin i$ cm. 

\begin{figure}
\centerline{\includegraphics[width=0.8\columnwidth]{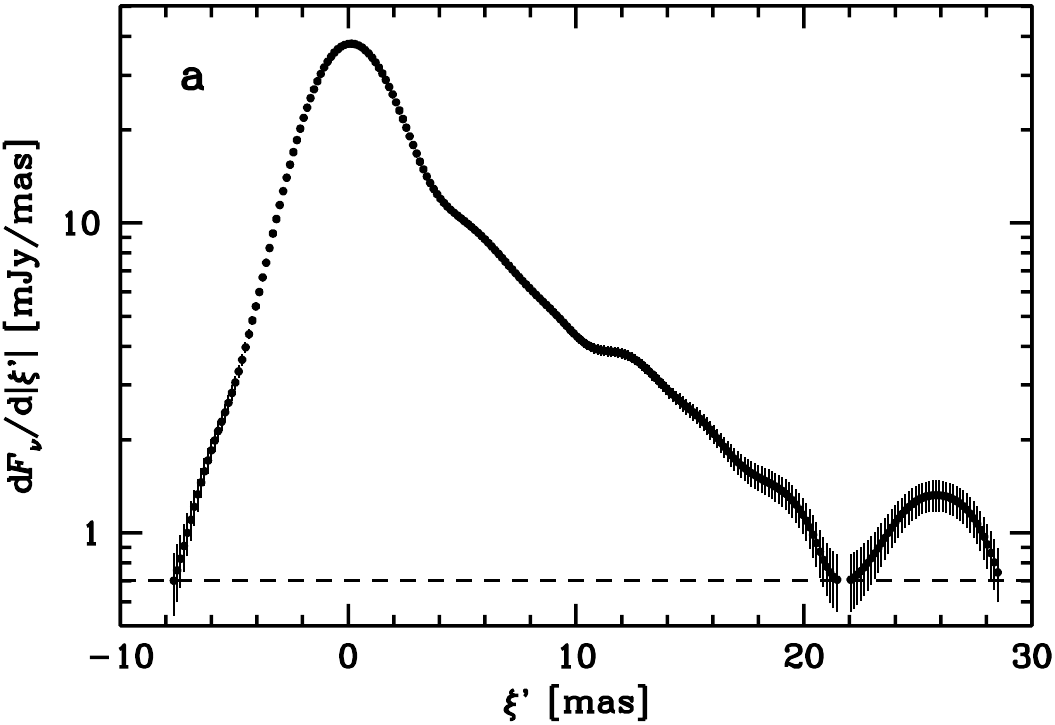}}
\centerline{\includegraphics[width=0.8\columnwidth]{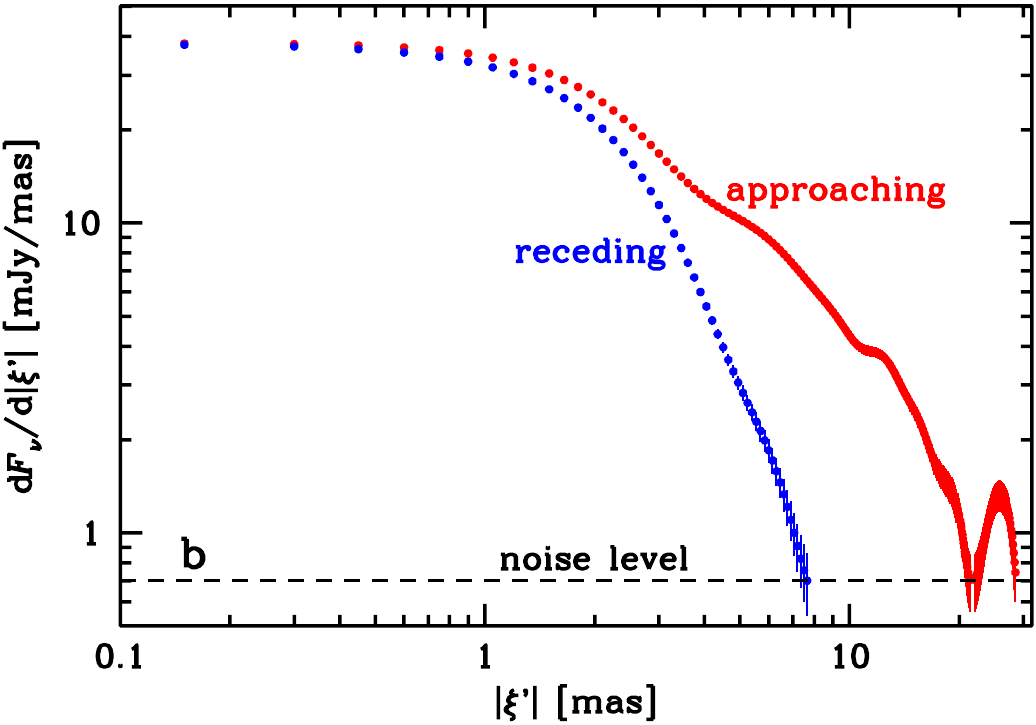}}
\caption{The emission profiles at 8.37 GHz of the jet and counterjet (\wood) adopting the zero point of \wood, determined by fitting a Gaussian to central data. The horizontal black line corresponds to the noise level. (a) The angular separation from the center of the core is plotted linearly. (b) The angular separation is in the logarithmic scale, with the red and blue points corresponding, respectively, to the approaching and receding jets.
}
 \label{data_plots}
 \end{figure}

We have recalculated the emission profile of the jet at 8.37 GHz, shown in \wood at points separated by 0.15 mas and including only the statistical errors. The systematic error, included in \wood but not here, was due to the global amplitude gain calibration being reliable only up to 10\%, which would only shift the entire profile up or down within that range. Since the jet is not resolved perpendicular to its length, the normalization corresponds almost exactly to the flux density per unit jet length in units of mas. We show the data in Figure \ref{data_plots} in the linear and log scales. In the bottom plot, we see that both the receding and approaching jets have almost the same flux densities up to $\xi'\approx 0.6$--1. While the shown data are convolved by the telescope restoring beam, we still expect the flux density of the approaching jet to be noticeably stronger (see Section \ref{model}). The lack of it could be partly due to the centroid image point of \wood being offset from that of the BH. 

Indeed, a core shift in \source was measured by \citet{Wood25} in an observation on 2023 September 20. The angular shift between the centroids of the 2.3 and 8.3 GHz compact-jet emissions was measured as $3.78\pm 0.15$ mas. Assuming the peak flux at a given frequency scales as $\nu^{-1}$ \citep{BK79}, we obtain the distance of the 8.3 GHz centroid from the BH of $\Delta\xi \approx 1.45\pm 0.05$ mas, which equals $8.7\times 10^{13}$ cm at the distance of 4 kpc. However, the jet studied by us had the integrated flux of $F_\nu=101\pm 10$ mJy (\wood), while the compact core observed on 2023 September 20 had $F_\nu=20\pm 2$ mJy \citep{Wood25}. The jet length scales $\propto F_\nu^{8/17}$ in the model of \citet{BK79}, see, e.g., equation (5) in \citet{Heinz06}, which would imply $\Delta\xi \approx 3.0$ mas at face value for the studied observation. A core shift increasing with the flux was seen for another BH XRB, MAXI J659--152 \citep{Paragi13}, see their figures 4 and 5. However, the two observations of \source were taken at different times and in different states, while the scaling assumes constancy of the jet parameters other than the luminosity. Here, we consider the range of $1.5\leq \Delta\xi/{\rm mas}\leq 3.0$ mas, corresponding to the core-BH distances at 4 kpc of (9--$18)\times 10^{13}$ cm. 

For comparison, the distance from the BH of the bulk of the 8.5 GHz emission in the hard state of the X-ray binary MAXI J1820+070, determined using the multi-frequency time-lag and spectral data of \citet{Tetarenko21} was $\approx\! 4\times 10^{13}$ cm, and the distance determined by the variability properties was about twice as large, see figure 7 in \citet{Zdziarski22a}. The photospheric distance at 8.4 GHz of Cyg X-1 was estimated by \citet{Heinz06} as $\approx 7\times 10^{13}$ cm (given here for the distance and the inclination of Cyg X-1 of \citealt{Miller-Jones21}). 

In our calculations, we also take into account the spectral measurement between 5.25 and 7.45 GHz of \citet{Miller-Jones23}, who obtained the spectral index of $\alpha=0.19\pm 0.07$. Although that measurement was taken four days before the VLBA observation, the measured spectral index is typical for the hard state of accreting BH systems, e.g., \citet{Tetarenko21}.

\section{The jet model}
\label{model}

Our model is based on that of \citet{BK79} and \citet{Konigl81}. We consider a conical jet with a radius $r(z)=\theta |z|$, where $\theta$ is the half-opening angle, constrained by the observations to $\theta\lesssim 0.5\degr$ (\wood). Negative values of $z$ correspond to the counterjet. Then, we assume a constant velocity, $\beta c$, with the corresponding bulk Lorentz factor denoted as $\Gamma$. The electrons in the jet, with the comoving power-law distribution, $N(\gamma,z)\propto \gamma^{-p}$, are accelerated and/or reaccelerated within Lorentz factors $\gamma_{\rm min}<\gamma< \gamma_{\rm max}$ and within the distances from the BH of $z_{\rm m}\leq z\leq z_{\rm M}$. Below $z_{\rm m}$, there is no emission, but that distance corresponds to a separation of $\ll$1 mas; hence we assume $z_{\rm m}=0$. Above $z_{\rm M}$, our model is no longer applicable. We adopt the assumption of \citet{BK79} that electron reacceleration compensates for the radiative and adiabatic losses. 

Details of our model are given in Appendix \ref{jet}. Our main result is that the spatial distribution of $F_\nu$ can be expressed as
\begin{equation}
\frac{{\rm d}F_\nu}{{\rm d}|\xi|}=A_1\delta^{\frac{1}{2}} |\xi|^{1+\frac{b}{2}}\! \left[1-\exp\!\left(\!-A_2\delta^{1+\frac{p}{2}} |\xi|^{-\frac{b p}{2}-b-1}\right) \right].\label{Fxi}
\end{equation}
We assume a value of $z_{\rm M}$ beyond the observed range. This formula gives the flux density and the synchrotron optical depth of Equations (\ref{tausyn}--\ref{Fz}) in terms of the separation in units of 1 mas, $\xi$, as defined by Equation (\ref{mas}). Here, $\delta$ is the Doppler factor, Equation (\ref{nu}), $b$ is the index of the power-law spatial dependence of the magnetic field strength, Equation (\ref{B}), and the constants $A_1$ [in units of erg/(s Hz mas)] and $A_2$ are given by Equations (\ref{A1}--\ref{A2}). The constants depend on $\beta$, $\theta$, $p$, $\nu$, $i$, $\gamma_{\rm min}$, $\gamma_{\rm max}$, $M_1$, $D$, the effective mass flow rate through the jet defined by Equation (\ref{mdot}), $\dot m_{\rm eff}$, and the reference magnetic field strength, $B_1$, Equation (\ref{B}).  

We then need to convolve ${\rm d}F_\nu/{\rm d}|\xi|$ with the Gaussian restoring beam of the telescope. Its distribution is given by
\begin{equation}
G(\xi)=\frac{\exp\frac{-\xi^2}{2 \sigma_{\rm t} ^2}}{\sqrt{2 \pi } \sigma_{\rm t} }.
\label{gauss}
\end{equation}
For the observation studied here, the FWHM of the beam equals 2.5 mas, which corresponds to $\sigma_{\rm t}=1.06$ mas. Then, the convolved distribution is
\begin{equation}
\frac{{\rm d}F_{\rm \nu,c}}{{\rm d}|\xi|}=\int_{-\infty}^{+\infty} G(\xi-x)\frac{{\rm d}F_\nu}{{\rm d}|x|} d x.
\label{conv}
\end{equation}
This effect causes the jet profile centered on the core within the FWHM of the beam to be approximately flat. We note that this Gaussian smoothing includes regions with negative and positive signs of $\xi$ for the jet and counterjet, respectively. In the case of different parameters for the jet and counterjet, which we find necessary for this source, we still calculate ${\rm d}F_{\rm \nu,c}/{\rm d}|\xi|$ for the jet and counterjet using their respective parameters for both positive and negative $\xi$. This is motivated by the jet/counterjet asymmetry being due to the time lags of evolving jets. 

\section{Results}
\label{results}

\begin{figure}
\centerline{\includegraphics[width=0.8\columnwidth]{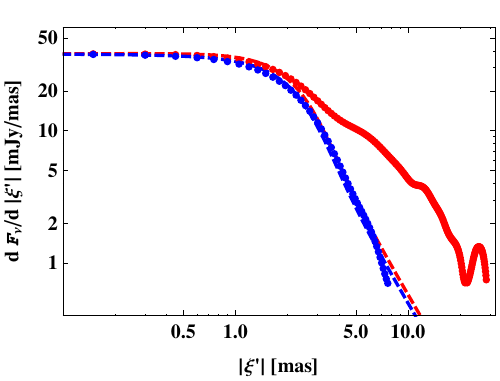}}
\caption{The emission profiles of the jet (red) and counterjet (blue) compared to our model assuming symmetric jets and $\Delta\xi=0$. We required the model to reproduce the range of $|\xi|\lesssim 1$, in which case the approaching jet model completely misses the profile at $|\xi|\gtrsim 2$. For clarity, the error bars are not shown.
}
 \label{no_offset}
 \end{figure}

\begin{figure}
\centerline{\includegraphics[width=0.8\columnwidth]{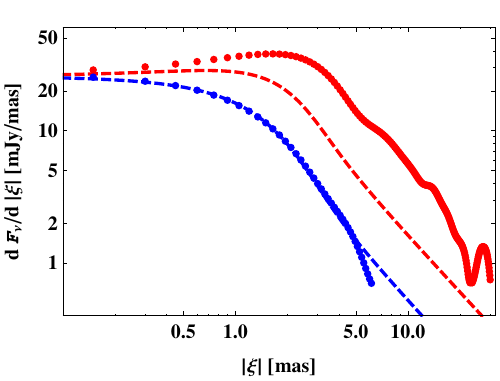}}
\caption{The emission profiles of the jet (red) and counterjet (blue) compared with our model assuming symmetric jets and $\Delta\xi=1.5$ mas, $i=45\degr$ and $\beta=0.35$. We found that any symmetric model fitting the counterjet misses the jet profile for any $\Delta\xi$.
}
 \label{offset15}
 \end{figure}

\begin{figure}
\centerline{\includegraphics[width=0.9\columnwidth]{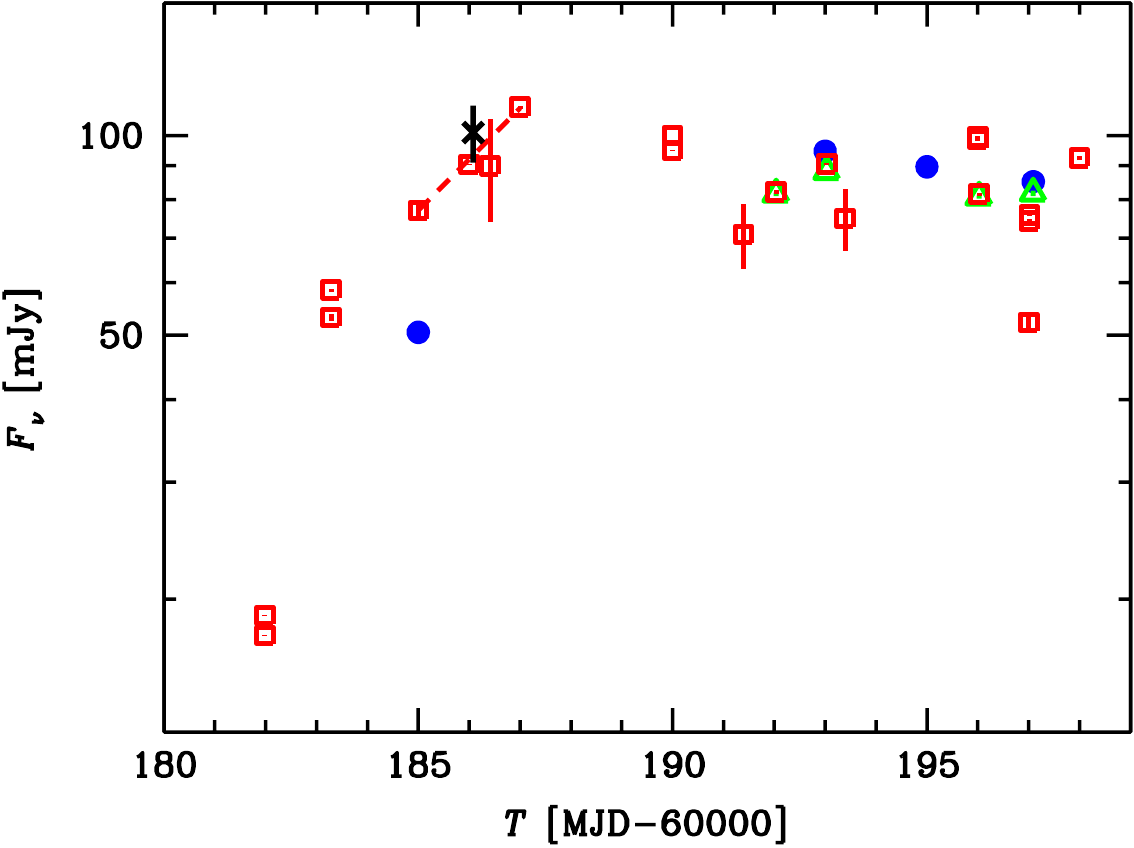}}
\caption{The radio light curve compiled from different measurements \citep{Hughes25}, including those reported in \wood. The VLBA observation is shown by the black cross with an error bar. The measurements at 5--9 GHz, 3 GHz, and 1.3--1.5 GHz are shown by red open squares, green open triangles, and blue filled circles, respectively. We see that the VLBA observation was done close to the end of the rising part of the light curve. 
}
 \label{lc}
 \end{figure}

\begin{figure}[t!]
\centerline{\includegraphics[width=1.0\columnwidth]{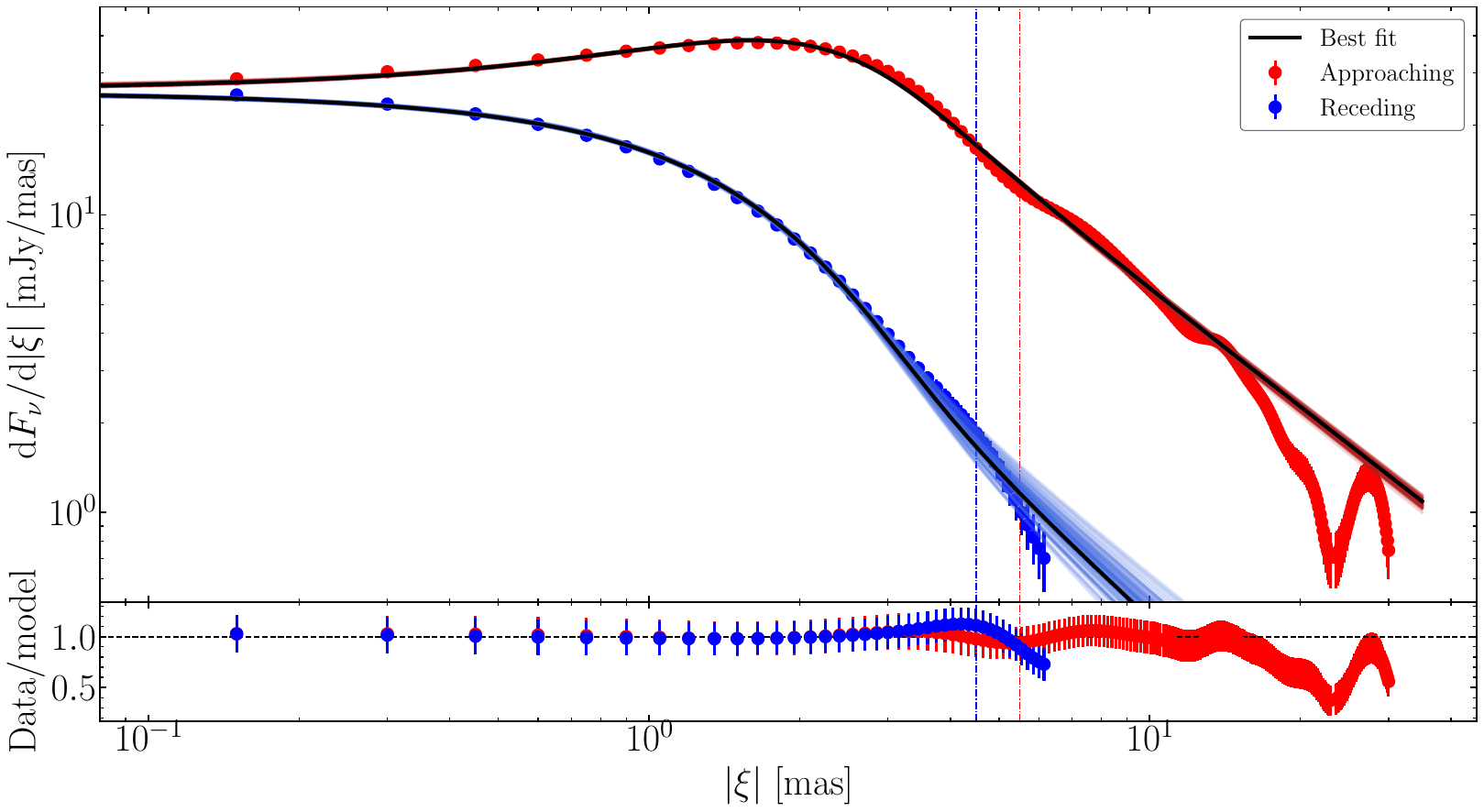}}
\centerline{\includegraphics[width=1.0\columnwidth]{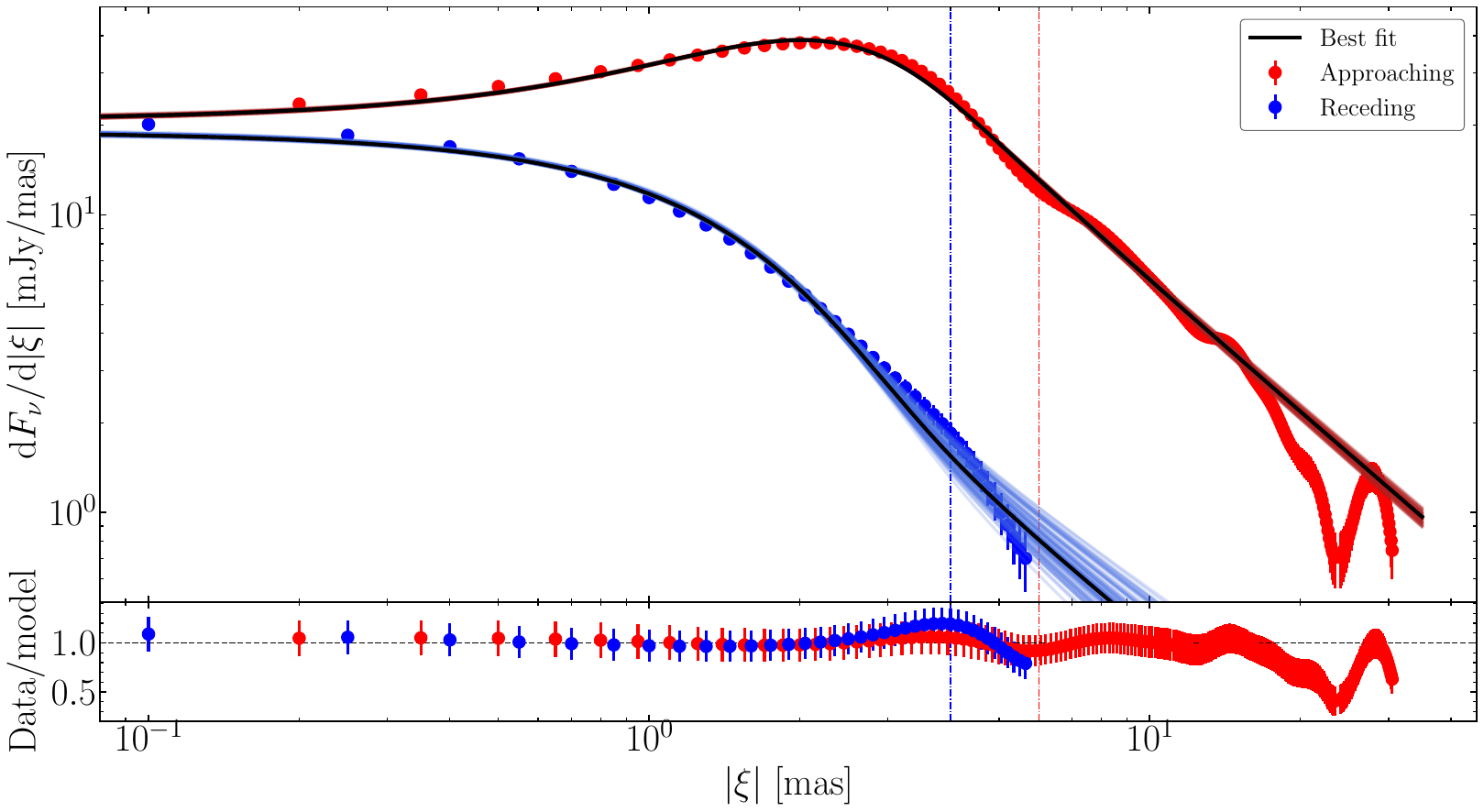}}
\caption{The profiles fitted to separations of $|\xi'| \leq 4$ and 6 mas for the jet and counterjet, respectively (shown by the vertical dot-dashed lines) for $\Delta\xi=1.5$ mas (top), and $\Delta\xi=2.0$ mas (bottom). We allow for different parameters of the jet and counterjet, see Table \ref{fits}. The flux profiles correspond to the final positions of the MCMC walkers in the model parameter space, and the red and blue shaded areas represent the total uncertainty on the fits.
}
 \label{offset_fit}
 \end{figure}

\begin{figure}
\centerline{\includegraphics[width=0.8\columnwidth]{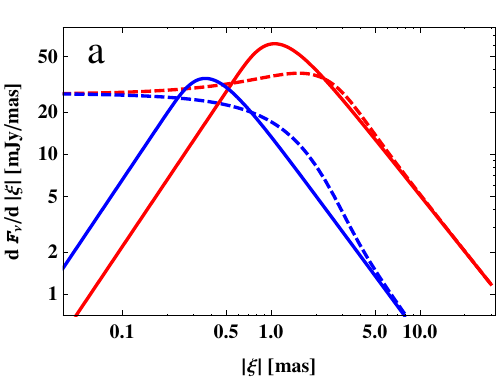}}
\centerline{\includegraphics[width=0.8\columnwidth]{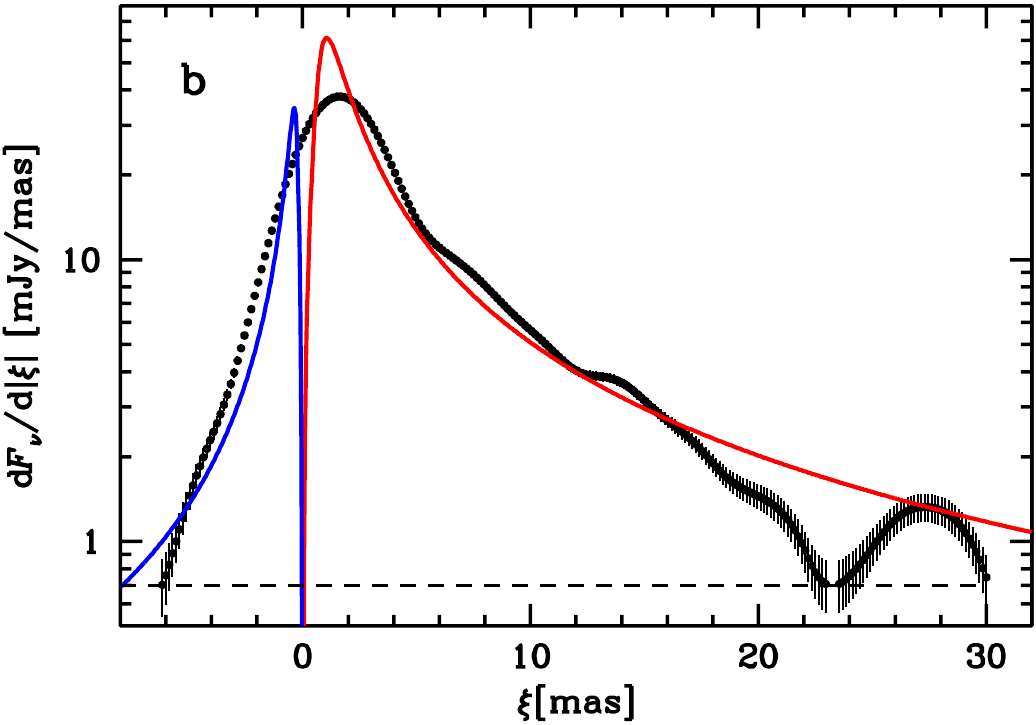}}
\caption{(a) The model emission profiles of the jet (red) and counterjet (blue) for $\Delta\xi=1.5$ mas, and with the same parameters as in Figure \ref{offset_fit}. The solid and dashed curves correspond to the emitted fluxes and those observed by the telescope, i.e., convolved with its restoring beam, respectively. (b) The emitted model fluxes (before the Gaussian smoothing) are compared with the data.
}
 \label{offset_m}
 \end{figure}

Here, we assume $b=1.17$, for which $\alpha\approx 0.18$--0.19 in a wide range of $p$, see Equation (\ref{alpha}). The value of $\alpha=0.19$ was measured by \citet{Miller-Jones23}; see Section \ref{data}. We assume the same velocity and inclination for both jets.

We first test a case with $\Delta\xi=0$. If we want to reproduce both of the profiles at $|\xi|\lesssim 2$, we need the ratio of $\delta$ of the jet to counterjet to be very close to unity (see Equation \ref{ratio}), and, then we miss altogether the profile of the approaching jet at $|\xi|\gtrsim 2$. We show an example of such a model in Figure \ref{no_offset}, at $\beta=0.1$ and $i=75\degr$. 

We next consider the case with $\Delta\xi=1.5$ mas, see Section \ref{data}, and hereafter assume $i=45\degr$; see Section \ref{parameters}. We show an example assuming the jet and counterjet to be symmetric in Figure \ref{offset15}, for $\beta=0.35$. The approaching jet has a much flatter profile than the receding one, and we found that no intrinsically symmetric model can (even roughly) reproduce both jets. 

Given that, we have looked at the time dependence of the radio emission close in time to the VLBA measurement. In Figure \ref{lc}, we see a significant flux increase on the time scale of a day before the studied observation. Given the observed sizes of the jets and the distance estimates, the time lag due to the light travel time of the receding jet behind the approaching one is also of the order of a day. While, in principle, BH jets may be intrinsically asymmetric, with one side being associated with a stronger magnetic field (e.g, \citealt{Wang92}), we instead suggest that the apparent asymmetry of the jet in \source is due to time-delay effects.

Thus, we consider models allowing for the jet and counterjet to have some parameters different. We have performed a Markov Chain Monte Carlo (MCMC) study of the parameter space. We used the package {\sc emcee} \citep{Foreman-Mackey13}. Every MCMC run was conducted using 80 walkers. For each run, after manual inspection, we consider that convergence is reached when the positions of the walkers in the parameter space are no longer significantly evolving. Once the chains have converged, the best fit result for each parameter is taken as the median of the one-dimensional posterior distribution obtained from the converged chains, while the $1\sigma$ uncertainties are reported as the difference between the median and the 16th percentile of the posterior (lower error bar), and the difference between the 84th percentile and the median (upper error bar). Flat priors were assumed for all parameters.

\setlength{\tabcolsep}{3pt}
\begin{table}[t!]
\centering
\caption{The parameters obtained from MCMC}
\vskip -0.4cm
\begin{tabular}{ccccccc}
\hline
$\Delta\xi$ & $i$ & $\beta$ & Jet & $p$ & $A_1$ & $A_2$
\\
\hline
1.5\,mas & 45\degr & $0.35^{+0.01}_{-0.01}$ & Appr. &  $1.30^{+0.01}_{-0.01}$ & $74^{+1}_{-1}$ & $0.94_{-0.04}^{+0.04}$
\\
&&& Rec. & $1.52^{+0.08}_{-0.08}$ & $260^{+30}_{-30}$ & $0.11_{-0.02}^{+0.02}$
\\
\hline
2.0\,mas & 45\degr & $0.36^{+0.01}_{-0.01}$ & Appr. & $1.45^{+0.02}_{-0.02}$ & $37^{+1}_{-1}$ & $2.6_{-0.1}^{+0.1}$
\\
&&& Rec. & $1.33^{+0.08}_{-0.07}$ & $280^{+40}_{-30}$ & $0.06_{-0.01}^{+0.01}$
\\
\hline
\end{tabular}
\tablecomments{$A_1$ is in units of 1 mJy/mas.}
\label{fits}
\end{table}

\setlength{\tabcolsep}{4pt}
\begin{table*}[t!]
\centering
\caption{The derived physical quantities corresponding to the median MCMC values}
\begin{tabular}{ccccccccccccc}
\hline
$\Delta\xi$ & $i$ & $\theta$ & Jet & $B_1$ & $\dot m_{\rm eff}$ & $P_{\rm i}$ & $P_B$ & $P_{\rm e}$ & $\beta_{\rm eq,1}$ & $\sigma_1$ &$\phi_{\rm BH}$ \\
mas & $\degr$ & $\degr$ && G &  & erg s$^{-1}$ & erg s$^{-1}$ & erg s$^{-1}$ & & \\
\hline
1.5 & 45 &0.5 & Appr. & 3.3 & 0.0026 & $1.4\times 10^{34}$ & $1.2\times 10^{34}$ &$2.8 \times 10^{31}$ & 0.0036 & 110 & 0.044
\\
&&& Rec. & 0.18 & 0.037 & $2.1\times 10^{35}$ & $3.5\times 10^{31}$ &$2.2 \times 10^{32}$ & 9.4 & 0.023 & 0.016
\\
\hline
2.0 & 45 & 0.5 & Appr. & 9.0 & 0.0010 & $5.6\times 10^{33}$ & $8.8\times 10^{34}$ &$7.2 \times 10^{30}$ & 0.00012 & 2200 & 0.12
\\
&&& Rec. & 0.15 & 0.027 & $1.5\times 10^{35}$ & $2.4\times 10^{31}$ &$1.5 \times 10^{32}$ & 9.3 & 0.022 & 0.014
\\
\hline
\end{tabular}
\tablecomments{The values of $B_1$ need to be multiplied by $(\theta/0.5\degr)^2$, and those of $\dot m_{\rm eff}$, by $(\theta/0.5\degr)^{-1-p}(D/4\,{\rm kpc})$. See Section \ref{results} for the assumed parameters and Appendix \ref{jet} for the formulae and scalings.
}
\label{derived}
\end{table*}

We fit the data only up to separations of $|\xi'| \leq 4$ mas for the jet since its flux strongly wiggles at larger $|\xi'|$, apparently due to interactions of the jet with clumps in the interstellar medium (not included in our model). This can be, in particular, the reason for the distinct rebrightening at $\xi'\sim 25$ mas. For the counterjet, we fit up to $|\xi'| \leq 6$ mas, since above we see a steeper slope, likely to due to a cessation of the electron reacceleration. The obtained best-fit profile at $\Delta\xi=1.5$ mas is shown in Figure \ref{offset_fit}(top), and the median parameters and $1\sigma$ uncertainties are given in Table \ref{fits}. We see that the obtained fit is relatively good. Figure \ref{offset_m}(a) shows both the theoretical spectra as observed by the telescope, i.e., convolved with its restoring beam (dashed curves; Equations \ref{gauss}--\ref{conv}), as well as the ones as emitted by the jets (solid curves). Figure \ref{offset_m}(a) compares the emitted spectra (before smoothing) with the data. Figure \ref{corner}(a) in Appendix \ref{corner_plot} shows the parameters with their uncertainties and their mutual correlations. We have also tested other values of $i$, and found that the jet velocity increases with the jet viewing angle, e.g., $\beta= 0.61_{-0.10}^{+0.07}$ at $i=65\degr$. Then, we fit the case with $\Delta\xi=2.0$ mas, see Table \ref{fits} and Figures \ref{offset_fit}(bottom) and \ref{corner}(b). This gives us a relatively good fit. However, we find the fit significantly worsens for higher values of $\Delta\xi$.

We then assume $\gamma_{\rm min}=10$, $\gamma_{\rm max}=10^4$, $D=4$ kpc,
$\theta=0.5\degr$, $f_{\rm e}(1+2 n_+/n_{\rm p})=1$, and $M_1\approx 8.2\msun$, which corresponds to the best-fit value of $f$, $M_2=0.2\msun$, and $i=45\degr$. The resulting derived quantities are shown in Table \ref{derived}, see equations in Appendix \ref{jet}; the subscript 1 denotes the respective value at 1 mas. The values of $\phi_{\rm BH}$ (Equation \ref{phi_BH}) are given for $a_*=0.9$, $\ell =0.5$, and $F_{\rm bol}= 5\times 10^{-7}$ erg cm$^{-2}$ s$^{-1}$ \citep{Liu24}. We see that the change of the conditions from the counterjet (seen at an earlier time) to the jet consists mostly of the jet becoming much more dominated by the magnetic field. Still, we find that both jets have magnetic fluxes much below those launched by magnetically arrested disks (MAD, at which $\phi_{\rm BH}\sim 50$, see Appendix \ref{jet}).

\section{Discussion}
\label{discussion}

Using our new analysis method of radio images, we have established that the jet and counterjet of \source observed by \wood are intrinsically asymmetric. The asymmetry consists mostly of the spatial power law slope of the jet being significantly flatter than that of the counterjet. Those observations took place during a rise of the radio light curve, which can then explain the asymmetry as being due to the jet evolution, together with the light-travel time lag of the counterjet to the jet. For example, the time lag between the counterjet and jet is $\sim$1 day for $D=4$ kpc, $i=30\degr$, and a 10 mas typical angular separation. 

We have modeled the two jets with our new model, which assumes steady-state solutions for each. We also assumed the same constant velocity for each of them. Given the observed time dependence of the radio flux, a more realistic model should take into account both the time dependence and the time lags, see, e.g., \citet{Miller-Jones04_time}. However, we have opted for the simple model, given our lack of knowledge of the details of the time dependence. Therefore, the obtained parameters should be considered as bearing substantial systematic uncertainties and tentative. In particular, systematic uncertainties can affect the jet power ratio, $P_{\rm i}/P_B$. This ratio is of the order of the square of the Alfv{\'e}n Mach number, $M_{\rm A}$, and we expect $M_{\rm A}^2 \gg 1$ on the considered spatial scales. On the other hand we have, for our default parameters, $P_{\rm i}/P_B\sim 1$ for the approaching jet for $\Delta\xi=1.5$ mas and even less for $\Delta\xi =2.0$ mas. Apart from possible systematic errors, we note that this ratio is $\propto f_{\rm e}^{-1}\theta^{-7-p}$, see Equations (\ref{B1}), (\ref{mdot1}), (\ref{P_i}), (\ref{P_B}), while $f_{\rm e}$ can be as low as $10^{-2}$ \citep{Sironi13} and we have only an upper limit on $\theta$, and $(\theta/0.5\degr)^{-7-p}$ can be easily large. Similarly, our finding of large increases of both the magnetic field strenth and the ratio of its energy density to the electron one, $\beta_{\rm eq}^{-1}$, between the observation times of the counterjet and jet, see Table \ref{derived}, should be considered as tentative. Since $\beta_{\rm eq}^{-1}\propto \theta^{7+p}$, the actual values of $\beta_{\rm eq}^{-1}$ can be significantly lower for a jet narrower than $0.5\degr$. 

Still, the adopted model, based on that of \citet{BK79}, fits the observed profiles relatively well, apart from some wiggles indicating local interactions with the surrounding medium or remnants of previous jet activity. We have found that the magnetic field strength increases with time and becomes much stronger for the jet than for the counterjet. We also find that the jet velocity is relatively low. 

An important result of our modelling is that the magnetic fields are relatively low. We find the (conserved) magnetic fluxes at large distances estimated by Equation (\ref{phi_BH}) imply the magnetic fluxes threading the BH to be much below those of jets launched by a MAD. Since $\phi_{\rm BH}\propto \theta^3$ (Equations \ref{B1}, \ref{phi_BH}), its values would become even lower if the jet is narrower than $0.5\degr$. This agrees with the conclusion of \citet{Barnier24}, obtained based on upper limits on the magnetic field in accretion flows of BH XRBs from the lack of substantial Faraday depolarization observed by IXPE. Also, it agrees with the conclusion of \citet{Z_Heinz24} that the power of compact jets in the hard state of BH LMXBs is much below that corresponding to the MAD.

We can also compare our values of the magnetic field strength with the predictions based on the core shift observed by \citet{Wood25} (at a different time). For this comparison, we use equation (7) of \citet{ZSPT15}, which yields the field strength at the distance corresponding to 1 mas, $p=2$, $D=4$ kpc, and $i=45\degr$ of $B_1\approx 8\beta_{\rm eq}^{-1/4}$ G, i.e., similar to the values found for the jet; see Table \ref{derived}.

Our new method works best for a steady state, in which both the jet and counterjet are steady, in which case light-travel time delays would play no role. The method can be applied to both jets launched from XRBs and AGNs. The method utilizes much more information (the jet spatial profiles) 
than the core-shift method, which is based on a single observable (the core displacement between two frequencies), and thus it is more powerful. The latter assumes a strict validity of the model of \citet{BK79} and requires assuming the value of the equipartition parameter, which is not the case for the new method.

We also found that the available information for \source favors relatively low distances, $\lesssim$5 kpc if $M_1\lesssim 13\msun$. This follows from our considerations in Section \ref{parameters}, where we show that the very high level of the maximum X-ray flux observed in the hard state would imply super-Eddington luminosities for $D\gtrsim 5$ kpc, while such luminosities have not been observed in a BH XRB in the hard spectral state before. 

An intriguing possibility is that the BH mass in \source is the highest one observed so far among LMXBs, $>13\msun$. This is hinted by both its very high peak X-ray flux and the large size of the compact X-ray jet, the largest observed as yet (\wood).

\section{Conclusions}
\label{conclusions}

Our main results are as follows.

We have developed a novel model for fitting the spatial structure of compact, continuous jets, applicable to both XRBs and AGNs. The model is based on that of \citet{BK79} and \citet{Konigl81}, in which the sum of partially synchrotron self-absorbed emission results in a flat radio spectrum, but it is in the physical space, tracking the fully self-absorbed emission followed by the optically thin one. Given that it utilizes much more information than the core-shift method, it is more powerful.

We have applied this model to the BH XRB \source. We have reproduced with it the main features of the spatial emission profiles of the hard-state compact jet. However, we have found the approaching jet to be significantly intrinsically stronger than the receding one. We attribute this difference to the light travel time effect, which is supported by the observed flux increase being on a similar time scale to the light travel time between the counterjet and the jet. Our model predicts the magnetic field strength along the jets and the mass flow rate due to the emitting relativistic electrons. We have found that while the magnetic field strength increases with time, the magnetic fluxes are still low and much below the MAD limit. 

The very high X-ray flux from the hard state and the very large size of its compact jet may suggest that the BH in \source may be the heaviest one among known LMXBs.  

\section*{Acknowledgements}

We acknowledge Ben Burridge, Chris Done, and James Miller-Jones for valuable discussions and the referee for their insightful suggestions. AAZ acknowledges support from the Polish National Science Center grants 2019/35/B/ST9/03944 and 2023/48/Q/ST9/00138. CMW acknowledges financial support from the Forrest Research Foundation Scholarship, the Jean-Pierre Macquart Scholarship, and the Australian Government Research Training Program Scholarship.

\appendix
\section{The model formalism}
\label{jet}

We follow the assumptions of \citet{BK79} and \citet{Konigl81}, but, in addition, we assume the conservation of the rest mass flux and provide explicit formulae for the relevant physical quantities. Also, we take into account the cosmological redshift, denoted here as $z_{\rm r}$. The jet length in the local observer's frame is denoted by $z$, 
\begin{equation}
z=\xi z_1,\,\, 
z_1\equiv \frac{1\,{\rm mas}\, D_A}{\sin i},\,\, 
\frac{{\rm d}z}{{\rm d}\xi}=z_1,\,\, D_A=\frac{D_L}{(1+z_{\rm r})^{2}},
\label{masz}
\end{equation}
and $D_L$ and $D_A$ are the luminosity and angular diameter distances, respectively. The jet is conical, with $r(z)=\theta |z|$. 

We assume the conservation of the rest mass flux, 
\begin{equation}
\dot M_{\rm j}=2 \pi \theta^2 z^2 \rho c \beta \Gamma,
\label{Mdot}
\end{equation}
where $\rho$ is the mass density in the comoving frame. This implies,
\begin{equation}
\rho(z)=\frac{\dot M_{\rm j}}{2\pi \theta^2 z^2 c \beta\Gamma}=\frac{2\dot m_{\rm j} G M_1 m_{\rm p}}{\theta^2 z^2 c^2 \beta\Gamma \sigma_{\rm T}},
\label{rho}
\end{equation}
where $\dot m_{\rm j}\equiv \dot M_{\rm j} c^2/L_{\rm Edd}$. A fraction, $f_{\rm e}\leq 1$, of the ion-related electrons is accelerated into a power law distribution, $\propto \gamma^{-p}$. In some simulations, the fraction of particles accelerated by internal shocks was found to be as low as $f_{\rm e}\sim 0.01$ \citep{Sironi13}. In addition, we allow for the presence of e$^\pm$ pairs with a density of $n_+$ in the power-law distribution, but we neglect their (typically small) contribution to the mass flux. We then define
\begin{equation}
\dot m_{\rm eff}\equiv f_{\rm e}(1+2n_+/n_{\rm p})\dot m_{\rm j},
\label{mdot}
\end{equation}
where $n_{\rm p}$ is the ion density. This formula gives the dimensionless accretion rate corresponding to the electron acceleration. Consequently,
\begin{equation}
N(\gamma, z)=\frac{2 G M_1 \dot m_{\rm eff} (p-1) \gamma^{-p}}{\theta^2 |z|^2 c^2 \beta\Gamma\sigma_{\rm T}(\gamma_{\rm min}^{1-p}-\gamma_{\rm max}^{1-p})}.
\label{N}
\end{equation}
Note that \citet{Konigl81} introduced a general dependence of $N(\gamma, z)\propto |z|^{-a}$, accounting for departures from the conservation of the electron distribution integrated across the jet, which can give $a\neq 2$. Here, we do not use it since we link $N$ with $\dot M_{\rm j}$, and possible departures are included in the $f_{\rm e}(1+2n_+/n_{\rm p})$ terms. Also, we assumed $p>1$. At $|z|<z_{\rm m}$, $N(\gamma,z)=0$. 

We assume a power-law distribution of the magnetic field strength, 
\begin{equation}
B(z)=B_1 (|z|/z_1)^{-b},
\label{B}
\end{equation}
where $B_1$ is the value of $B$, respectively, at the reference distance, $z_1$, see Equation (\ref{mas}), and $b=1$ for conserved energy flux of toroidal magnetic field. The spectral index, $\alpha$, of the partially self-absorbed part of the spectrum is then a function of $a$, $b$, and $p$, 
\begin{equation}
\alpha=\frac{5 a+3 b+2(b-1)p-13}{2a-2+b(p+2)}
\label{alpha}
\end{equation}
as given by equation (A10) of \citet{Zdziarski19b}. Values of $b>1$ correspond to some decay of magnetic field energy flux with the distance, due to, e.g., magnetic reconnection or conversion of the magnetic energy into either bulk or electron acceleration.

We then adopt a delta-function approximation for the observed synchrotron photon frequency, $\nu$,
\begin{equation}
\frac{h \nu}{m_{\rm e}c^2}\approx \frac{B}{B_{\rm cr}}\frac{\delta\gamma^2}{1+z_{\rm r}},\quad B_{\rm cr}\equiv \frac{2\pi m_{\rm e}^2 c^3}{e h},\quad
\delta=\frac{(1-\beta^2)^{1/2}}{1-\beta\cos i},
\label{nu}
\end{equation}
where $B_{\rm cr}$ is the critical magnetic field strength, $e$ is the electron charge, $m_{\rm e}$ is the electron mass, $h$ is the Planck constant, $\delta$ is the Doppler factor, and $i\rightarrow i+180\degr$ for the counterjet. Note that $\nu$ is the observed frequency, which transforms to the comoving one via the $\delta/(1+z_{\rm r})$ factor, which is different for the jet and counterjet. 

We use a version of the model of \citet{BK79} as developed in Appendix A of \citet{Zdziarski19b}. In particular, we use their equations (A4) and (A5), the latter following from integration of the source function over the jet projected area. The optical depth to synchrotron self-absorption is,
\begin{align}
&\tau(\nu,z)= \label{tausyn}\\
&\frac{2\pi C_2(p) (p-1) [\delta c^2 B(z)/B_{\rm cr} ]^{1+\frac{p}{2}} G M_1\dot m_{\rm eff} }
{\alpha_{\rm f} \theta |z| (\gamma_{\rm min}^{1-p}-\gamma_{\rm max}^{1-p})[h\nu(1+z_{\rm r})/m_{\rm e}]^{2+\frac{p}{2}}\beta\Gamma \sin i},\nonumber
\end{align}
where $\alpha_{\rm f}$ is the fine-structure constant. The distribution of the flux per unit distance can be expressed as
\begin{align}
&\frac{{\rm d}F_\nu}{{\rm d}|z|}=\frac{(1+z_{\rm r})^{7/2} C_1(p) \pi (m_{\rm e} B_{\rm cr} h\delta)^{1/2} \nu^{5/2}\theta |z|\sin i }{6 C_2(p) B(z)^{1/2}c D_L^2} \times \nonumber\\
&\left[1-\exp-\tau(\nu,z) \right],\quad z_{\rm m}\leq |z|\leq z_{\rm M},
\label{Fz}
\end{align}
where ${\rm d}F_\nu/{\rm d}z=0$ for $|z|<z_{\rm m}$. Above, $C_1(p)$ and $C_2(p)$ are constants of the order of unity, given, e.g., by equations (8--9) of \citet{Zdziarski22a}. For steady-state, symmetric, jets, Equations (\ref{tausyn}--\ref{Fz}) imply the jet-to-counterjet flux ratios of
\begin{equation}
\left(\frac{1+\beta\cos i}{1-\beta\cos i}\right)^{1/2},\quad \left(\frac{1+\beta\cos i}{1-\beta\cos i}\right)^{(3+p)/2},
\label{ratio}
\end{equation}
in the optically thick and optically thin jet parts, respectively. 

Next, we express ${\rm d}F_\nu/{\rm d}z$ and $\tau$ of Equations (\ref{Fz}) and (\ref{tausyn}), respectively, in terms of $\xi$, 
\begin{equation}
\frac{{\rm d}F_\nu}{{\rm d}|\xi|}=A_1\delta^{\frac{1}{2}} |\xi|^{1+\frac{b}{2}}\! \left[1-\exp\!\left(\!-A_2\delta^{1+\frac{p}{2}} |\xi|^{-\frac{b p}{2} - b + 1 - a}\right) \right],\label{Fxi2}
\end{equation}
where we now included the dependence on $a$ for completeness, while Equation (\ref{Fxi}) is given for $a=2$. In the optically thick and optically thin regimes, we have the asymptotic dependencies,
\begin{equation}
\frac{{\rm d}F_\nu}{{\rm d}|\xi|}\approx \begin{cases}A_1\delta^{\frac{1}{2}} |\xi|^{1+\frac{b}{2}}, & \tau\gg 1;\cr
A_1 A_2\delta^{\frac{3+p}{2}} |\xi|^{-\frac{b p}{2} - b + 1 - a},  & \tau\ll 1, \cr\end{cases}
\label{asymptotic}
\end{equation}
and $\tau=1$ is at
\begin{equation}
|\xi|=\left(A_2 \delta^{\frac{2+p}{2}}\right)^\frac{2}{bp+2b-2+2a}.
\label{tau1}
\end{equation}
The constants are
\begin{align}
&A_1=\frac{C_1(p) \pi (m_{\rm e} h\theta B_{\rm cr})^{1/2} \nu^{5/2} (1\,{\rm mas})^2}{6 C_2(p)(1+z_{\rm r})^{1/2} B_1^{1/2}c \sin i}, \label{A1}\\
&A_2=\label{A2}\\ 
&\frac{2\pi C_2(p)(p-1) (c^2 B_1/B_{\rm cr})^{1+\frac{p}{2}}G M_1 \dot m_{\rm eff}}{\alpha_{\rm f}\theta (1+z_{\rm r})^{\frac{p}{2}}(1\,{\rm mas})D_L (h\nu/m_{\rm e})^{2+\frac{p}{2}}(\gamma_{\rm min}^{1-p}-\gamma_{\rm max}^{1-p})\beta\Gamma}\nonumber.
\end{align}
The flux density per unit angular separation in the optically thick part of the jet, $\propto A_1$, is independent of the distance for $z_{\rm r}\ll 1$. This follows from the area of a 1 mas strip of the jet having the physical area being $\propto D_A^2$, while the observed flux density is $\propto D_L^{-2}$. Also, it is independent of $\dot m_{\rm j}$, which follows from $\tau\gg 1$ and the nonthermal synchrotron source function being independent of the electron distribution, see equation (17) of \citet{ZLS12}.

The emitted flux density, Equation (\ref{Fxi2}), needs to be then convolved with the restoring beam of the telescope, Equation (\ref{gauss}, yielding the theoretical observed flux density, Equation (\ref{conv}). We then fit it to the data. For an assumed value of $b$, this yields $A_1$, $A_2$, $\delta(i,\beta)$, and $p$ for each jet. We then assume values of $i$ (yielding $\beta$ for the fitted $\delta$), $\theta$, $D$, $i$, $M_1$, $\gamma_{\rm min}$ and $\gamma_{\rm max}$. Then, Equation (\ref{A1}) yields $B$ at $z$ corresponding to $\xi=1$,
\begin{align}
&B_1=\frac{[C_1(p) \pi \theta]^2 m_{\rm e}h B_{\rm cr} \nu^5 (1\,{\rm mas})^4 }{[6 A_1 C_2(p) c \sin i]^2(1+z_{\rm r})} \approx 0.27\,{\rm G}\times
\label{B1}\\
&\left(\frac{C_1(p)}{C_2(p)} \frac{\theta}{1\degr} \frac{\sin 45\degr}{\sin i} \frac{1\,{\rm mJy/mas}}{A_1} \right)^2 \!\left(\frac{\nu}{1\,{\rm GHz}}\right)^5\!\! \frac{1}{1+z_{\rm r}}.
\nonumber
\end{align}
The value of $B_1$ can be compared with that from core shift, if available. With the above, we can solve Equation (\ref{A2}) for $\dot m_{\rm eff}$,
\begin{align}
&\dot m_{\rm eff}=\label{mdot1}\\
&\frac{(6 A_1 \sin i)^{2+p}\!A_2 \alpha_{\rm f}\beta\Gamma D_L h C_2(p)^2 (1\!+\!z_{\rm r})^{1+p} (\gamma_{\rm min}^{1-p}\!-\!\gamma_{\rm max}^{1-p})}{2 C_1(p)^{2+p} (\pi m_{\rm e})^{3+p} (p-1) (1\,{\rm mas}\,\nu)^{3+2p}\theta^{1+p} G M_1 }.\nonumber
\end{align}

The component of the jet power in the bulk motion of ions is $\dot M_{\rm j}c^2 (\Gamma-1)$, which can be written as
\begin{equation}
P_{\rm i}=\frac{\dot m_{\rm eff} G M_1 m_{\rm p}c(\Gamma-1)}{f_{\rm e}(1+2n_+/n_{\rm p}) \sigma_{\rm T}}.
\label{P_i}
\end{equation}
For a toroidal magnetic field (dominating at large distances), its power is
\begin{equation}
P_{B}=(B^2/2)c \beta\Gamma^2 (z\theta)^2.
\label{P_B}
\end{equation}
We also define the equipartition parameter as the ratio of the electron to magnetic field energy densities, 
\begin{equation}
\beta_{\rm eq}\equiv \frac{u_{\rm e}}{B^2/8\pi},
\label{beta_eq}
\end{equation}
where $u_{\rm e}$ is the electron energy density. For $p< 2$, it depends sensitively on the maximum electron Lorentz factor. Then, the jet power in the electrons and positrons is
\begin{equation}
P_{\rm e}=(2\beta_{\rm eq}/3)P_B.
\label{P_e}
\end{equation}
We also define the magnetization parameter,
\begin{equation}
\sigma\equiv \frac{{B}^2}{4\pi \rho c^2},
\label{sigma}
\end{equation}
where $\rho$ is the mass density at $z$, see Equations (\ref{rho}). We note that $\rho$ depends on $\dot m_{\rm j}$, while we estimate only $\dot m_{\rm eff}$, see Equation (\ref{mdot}). Thus, an estimate of $\sigma$ based on $\dot m_{\rm eff}$ needs to be multiplied by $f_{\rm e}(1+2n_+/n_{\rm p})$.

Finally, we consider the magnetic flux. The flux, $\Phi_{\rm BH}$, of the poloidal magnetic field threading the BH on one hemisphere is conserved, and it equals that of the jet at large distances, $\Phi_{\rm j}$. In jet models with differential collimation of poloidal magnetic surfaces \citep{Tchekhovskoy09}, the ratio of the jet-frame poloidal-to-toroidal magnetic fields averaged over the jet cross section is,
\begin{equation}
\frac{B_{\rm p}(z)}{B(z)}\approx \frac{2^\frac{3}{2} r_{\rm H}\beta\Gamma}{a_* \ell r}\left(\frac{1+\sigma}{\sigma}\right)^\frac{1}{2},
\label{Bratio}
\end{equation}
where $r_{\rm H}$ is the BH horizon radius, $a_*$ is the dimensionless spin parameter, $\ell\approx 0.5$ is the ratio of the magnetic field rotation frequency to that of the BH at the horizon, and the conservation of the magnetic energy flux was assumed (see \citealt{ZSPT15}). Then,
\begin{equation}
\Phi_{\rm j}\equiv \pi B_{\rm p}(z) r(z)^2=\Phi_{\rm BH}\equiv \phi_{\rm BH}(\dot M_{\rm accr} c)^\frac{1}{2} r_{\rm g},
\label{Phi_BH}
\end{equation}
where $\dot M_{\rm accr}$ is the mass accretion rate and $\phi_{\rm BH} \lesssim 70$ is a scaling factor \citep{Tchekhovskoy11, Davis20}, where values close to the upper limit correspond to jets launched by magnetically-arrested disks (MAD; \citealt{BK74, Narayan03}). We have $\dot M_{\rm accr}c^2=4 \pi D^2 F_{\rm bol}/\eta$, where $\eta$ is the accretion efficiency, which, for simplicity, we take as $\eta(a_*)$ of thin disks \citep{Bardeen72}. This gives us\footnote{We notice that the formula for $\Phi_{\rm j}$ used in Equation (\ref{phi_BH}) is different from equation (33) in \citet{Zdziarski22a}. That equation was based on the derivation in \citet{Zamaninasab14}, which assumed the jet opening angle as $\theta\approx s\sigma^{1/2}/\Gamma$ \citep{Tchekhovskoy09}, where $s\lesssim 1$. Here we use the observationally constrained $\theta$, which is $\ll 1/\Gamma$, as is generally the case in X-ray binaries \citep{Miller-Jones06}. Therefore, the magnetic flux obtained in \citet{Zdziarski22a} for MAXI J1820+070 was substantially overestimated. }
\begin{align}
&\phi_{\rm BH}=\label{phi_BH}\\
&\frac{[2\pi \eta(a_*) c(1+\sigma)]^{1/2}(1+\sqrt{1-a_*^2}) (1\,{\rm mas})\beta\Gamma\theta B(\xi)\xi}{\ell a_* (\sigma F_{\rm bol})^{1/2}\sin i }.
\nonumber
\end{align}
We note that it is independent of $D$, which is due to $F_{\rm bol}^{1/2}\propto D^{-1}$ as well as $B(\xi)\propto D^{-1}$ at the assumed $b=1$.

\section{Comparison with other models}
\label{HJ88}

We first compare our model (which follows \citealt{BK79}) with the relatively similar model of \citet{Hjellming88}, who also studied the spatial flux distribution of conical jets. \citet{BK79} assumed that the jet's relativistic electrons have their energy conserved, implying that $N(\gamma, z)\propto z^{-2}$. The neglect of the adiabatic losses is justified by the electron energy remaining constant in a conical jet with constant speed \citep{Potter12, Potter15}; see also the discussion and references in \citet{Zdziarski19b}. When the energy flux of the toroidal magnetic field is conserved, $B(z)\propto z^{-1}$, the energy spectral index of the partially self-absorbed synchrotron radiation is $\alpha=0$, independent of the electron power law index, $p$. Similar values of $\alpha$ are indeed seen in both AGNs and BH XRBs \citep{Yuan18, Fender00}. \citet{Konigl81} considered then generalized dependencies, $N(\gamma, z)\propto z^{-a}$, $B(z)\propto z^{-b}$.

On the other hand, \citet{Hjellming88} assumed that the relativistic electrons after an initial acceleration undergo adiabatic losses. For $b=1$, which they assumed, this gives
\begin{equation}
a=\frac{2p+4}{3}, \quad \alpha=\frac{10(p-1)}{8+7 p},
\label{alphaHJ}
\end{equation}
which gives harder spectra than observed for typical values of $p$, e.g., $\alpha\approx 0.45$ for $p=2$. \citet{Hjellming88} also took into account a jet precession, and their geometry included an empty central conical spine. However, their formulae did not give the flux normalization expressed by the physical jet parameters.

A related model is that of \citet{Paragi13}. While in its description they invoked the model of \citet{Hjellming88}, they followed the approach of \citet{Konigl81} by using arbitrary values of $a$ and $b$. They provided formulae for ${\rm d}F_\nu/{\rm d}z$ and $\tau$ compatible with ours, except that they did not express the values of their constants in terms of physical quantities (as done by us in Appendix \ref{jet}). Also, their formula for ${\rm d}F_\nu/{\rm d}z$ included a correction factor, which, however, was derived for the original model of \citet{Hjellming88}, and it is not applicable for their formulation. 

\section{The values and correlations of the jet parameters}
\label{corner_plot}

We show here the correlations between the parameters as calculated with the MCMC, see also Table \ref{fits}.

\begin{figure*}
\centerline{\includegraphics[width=0.8\textwidth]{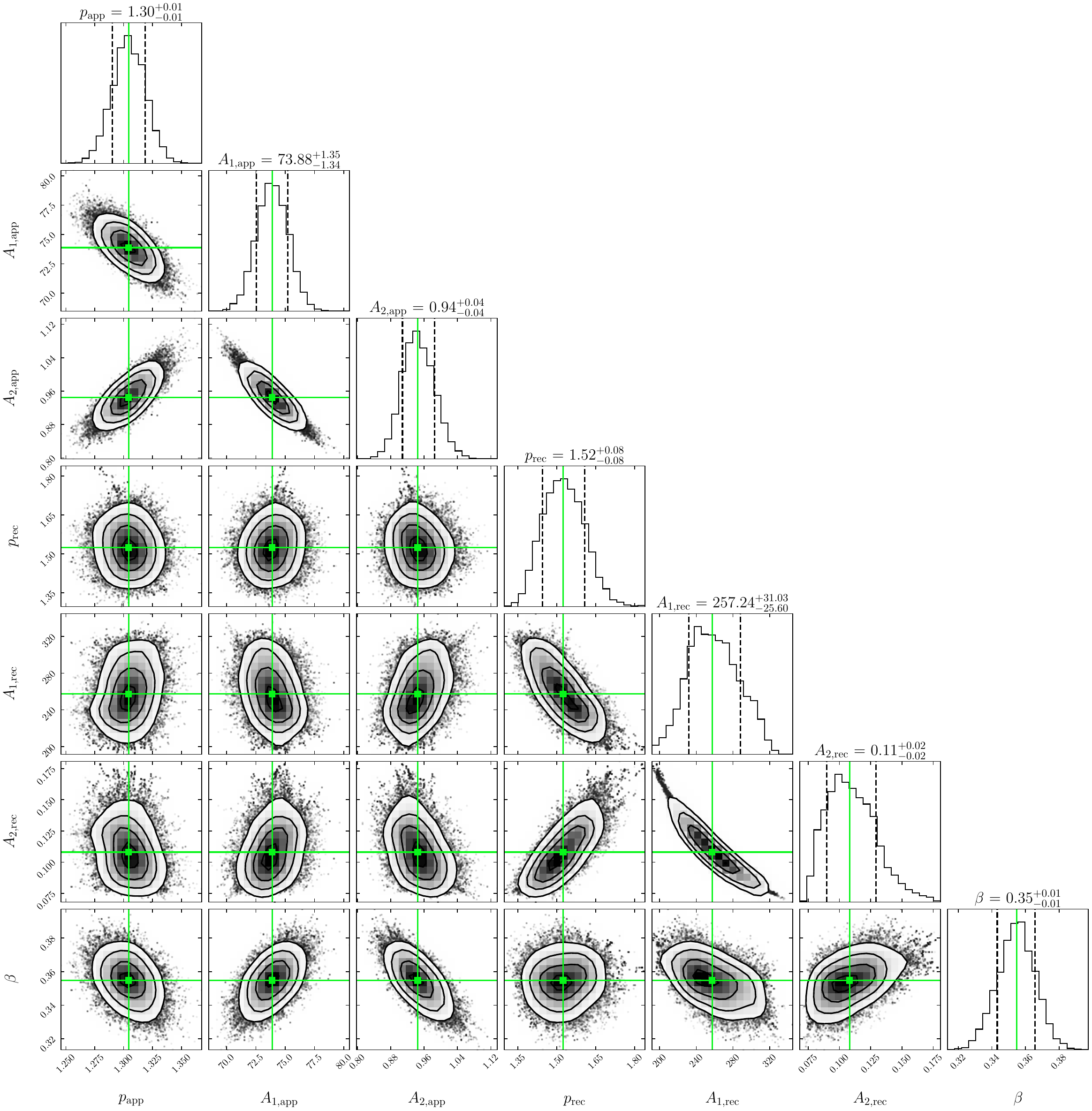}}
\caption{Correlations between the main parameters calculated by MCMC. The median values and the $\approx\! 1\sigma$ uncertainties are shown by the middle and surrounding dashed lines, respectively. The corresponding numerical values are given by the posterior distributions. (a) The fit with $\Delta\xi=1.5$ mas. 
}
 \label{corner}
 \end{figure*}

\setcounter{figure}{7}
\begin{figure*}
\centerline{\includegraphics[width=0.8\textwidth]{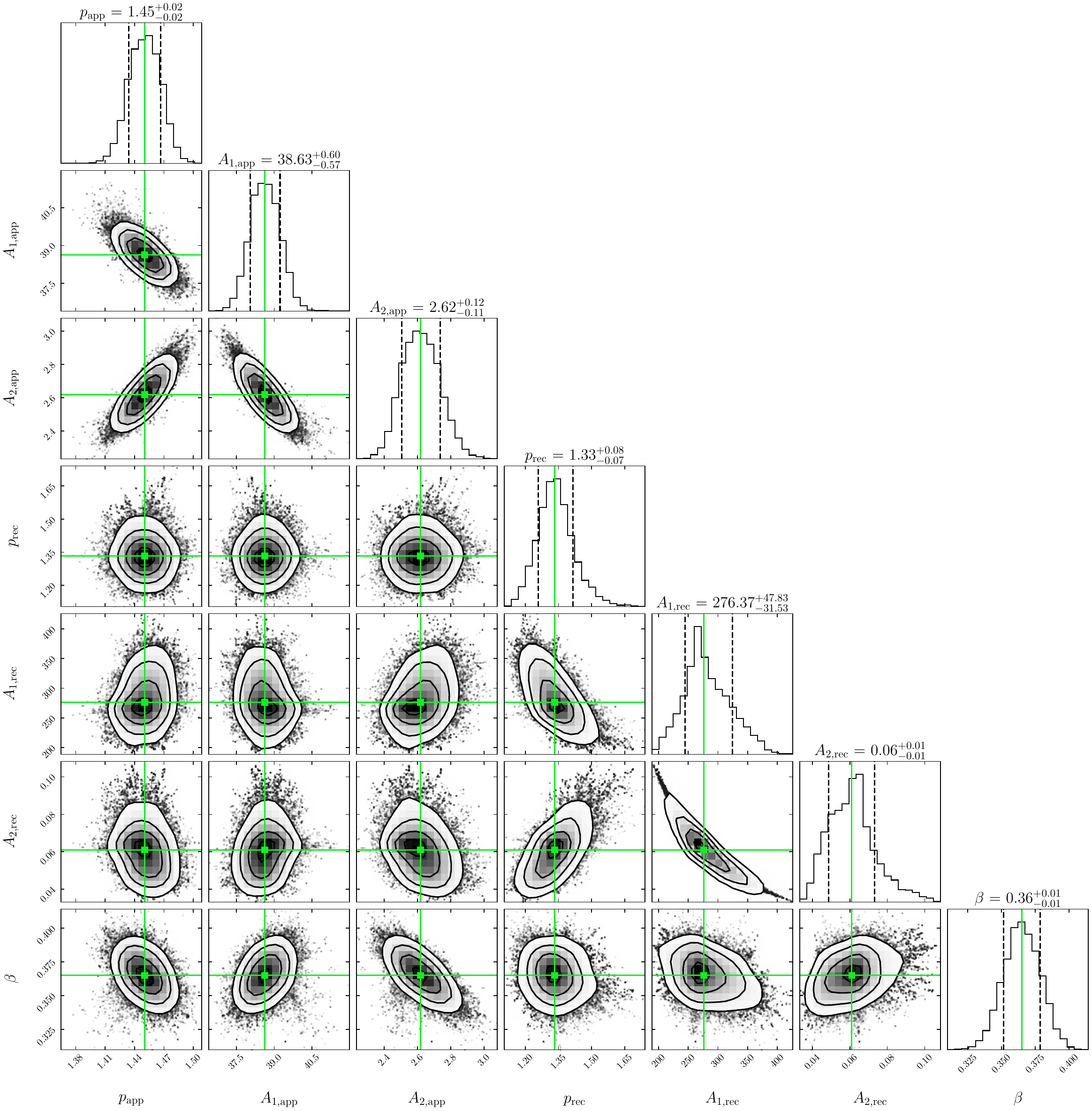}}
\caption{(b) The fit with $\Delta\xi=2.0$ mas .  
}
 \end{figure*}

\bibliographystyle{../aasjournal}
\bibliography{../allbib}{}

\end{document}